\documentstyle{article}
\catcode`\@ =11
\@addtoreset{equation}{section}

\catcode`\@ =12
\newtheorem{Theorem}{Theorem}[section]
\newtheorem{Proposition}{Proposition}[section]

\newtheorem{Lemma}{Lemma}[section]

\title{On some hydrodynamical aspects of quantum mechanics}

\author{
Mauro SPERA \\Dipartimento di Informatica, Universit\`a di Verona \\
Ca' Vignal 2, Strada le Grazie 15, 37134 Verona, Italia \\
e-mail: mauro.spera@univr.it}
 
\date{3rd February 2009}

\begin{document}
\textwidth=125mm
\textheight=185mm
\parindent=8mm
\frenchspacing
\maketitle
\begin{abstract}

In this note we first set up an analogy between spin and vorticity
of a perfect 2d-fluid flow, based on the Borel-Weil contruction
of the irreducible unitary representations of $SU(2)$, and looking at the
Madelung-Bohm velocity attached to the ensuing spin wave functions.
We also show that, in the framework of finite dimensional geometric quantum
mechanics, the Schr\"odinger velocity field on projective Hilbert space is
divergence-free (being Killing with respect to the Fubini-Study metric) and
fulfils the stationary Euler equation, with pressure proportional 
to the Hamiltonian uncertainty (squared).
We explicitly compute the pressure gradient of this ``Schr\"odinger fluid" and determine its critical points.
Its vorticity is also calculated and shown to depend on the spacings 
of the energy levels.
These results follow from hydrodynamical properties of Killing vector fields valid in any 
(finite dimensional) Riemannian manifold, of possible independent interest.

\par
\end{abstract}
\null

\section{Introduction}

The present note can be viewed as a follow-up of \cite{Be-Sa-Spe} and \cite{Ben-Spe} in that it explores geometric and
more generally ``classical" features of the {\it standard} quantum mechanical formalism, 
in the hope of shedding
some light on delicate conceptual issues, such as entanglement or quantum measurement 
(see the above references), or, at least, to get an intuitive grip on traditionally elusive topics.
So we first set up an analogy between spin and vorticity
of a perfect 2d-fluid flow, based on the Borel-Weil construction
of the irreducible unitary representations of $SU(2)$, and looking at the
(Madelung-Bohm) velocity attached to the ensuing spin wave functions.
This is motivated by the algebro-geometric approach to 2d-superfluids devised in \cite{Pe-Spe}
(see e.g. \cite{Gue} for physical background).
The point is that, unlike Borel-Weil, we have a {\it configuration} space interpretation of spin,
whereby putting standard and internal degrees of freedom on an equal footing. 
The  vortex strength interpretation of spin is consistent with the 
fact that vorticity is related to angular momentum, in the superfluid context (see Section 2, Theorem 2.1). The above spin wave functions are also interpreted
``semiclassically", in a suitable technical sense.\par
In Section 4, pursuing a hydrodynamical thread, we also show that, in the framework of finite dimensional geometric quantum
mechanics, the Schr\"odinger velocity field on projective Hilbert space is
divergence-free (being Killing with respect to the Fubini-Study metric) and
fulfils the stationary Euler equation, with the pressure being proportional to the Hamiltonian uncertainty (dispersion) - squared.
We explicitly compute the pressure gradient of this ``Schr\"odinger fluid" and determine its critical points. 
In particular, the energy eigenstates appear as the minimal (i.e. zero) pressure states,
an interpretation that could be relevant in the context of quantum measurement (collapse of the wave function).
The vorticity of the fluid is determined as well and shown to depend on the spacings of the energy levels.
These results are collected in Theorem 4.1 and
follow directly from (possibly new, or at least differently formulated)
``hydrodynamical" properties of Killing vector fields - of possible independent interest - valid in any 
(finite dimensional) Riemannian manifold, which we discuss in detail in Section 3
(see Theorem 3.1). It is perhaps useful to note at this point that, although complex projective spaces are dealt with in Sections 2 (the special case
${\bf P}^1 \equiv {\bf P}({\bf C}^2) \equiv S^2 $) and 4, (the general finite dimensional case), their actual roles in their respective contexts are completely different. The paper ends with 
some final comments and outlook.\par
 
\section{Spin = Vorticity}

We begin by recalling that the (unitary) $SU(2)$-representation of spin $s$ can be realized on the
$(2s+1)$-dimensional complex vector space consisting of all homogeneous
complex polynomials of degree $2s$ in $z_0$ and $z_1$ - homogeneous coordinates
on the Riemann sphere $S^2 \cong {\bf P} ( {\bf C}^2) \equiv {\bf P}^1$, 
with inhomogeneous coordinate $\zeta = {{z_1} \over {z_0}}$ (or the reciprocal) - 
whereupon $SU(2)$ acts via M\"obius transformations.
This is the simplest instance of the Borel-Weil construction of unitary representation
of simple Lie groups, and can be phrased in the language of (K\"ahlerian) geometric quantization,
see e.g. \cite{Ga-Spe} and \cite{Per}, Ch.4, for details, and also \cite{Tyu}, \cite{Wo} for general background);
here we just notice that the spin wave functions correspond to the holomorphic sections of
the $2s$-th tensor power ${\cal O}(2s) = {\cal O}(1)^{\otimes 2s}$
 of the hyperplane section bundle
${\cal O}(1) \rightarrow {\bf P}^1$, dual to the tautological
bundle. This is of course
in accordance with the fact that a particle of spin $s$ corresponds to a symmetric spinor of rank $2s$
(see \cite{LL}, Ch. VIII). 
In the fundamental representation, a spin Hamiltonian generates infinitesimal rotations around an axis
connecting the two eigenstates (see also Section 4, and \cite{Be-Sa-Spe},
\cite{Ben-Spe}, 
for a fairly general geometric picture
of Schr\"odinger's Hamiltonians).
A spin wave function ultimately becomes a polynomial $\chi = \chi(\zeta)$ of degree $2s$, and can be viewed as
a meromorphic function on $S^2$ (i.e. a rational
function, in this case), with a pole of order $2s$ at infinity (for a meromorphic function on a
compact Riemann surface one has  number of zeros =
number of poles, both counted according to their multiplicity (see \cite{G} or \cite{N}).
The functions $ \chi_k := \zeta^k$, $k=0,...,2s$ become, after suitable normalization, an
orthonormal basis for the spin space. 
The spin operator ``$S_z$" reads, in the above basis, 
$S_z {\chi}_k = (k - s) \chi_k$.
Therefore, $S^2 \cong  {\bf P}^1$ is the classical {\it phase} space attached to spin (cf. \cite{Per}).
\par
The above arrangement matches (in the genus zero case) exactly the algebro-geometric description of superfluids
(more precisely, of their order parameters) devised, e.g. in
\cite{Pe-Spe}. Pursuing the analogy in detail we introduce, for a spin wave function
$\chi = \chi(\zeta) = \prod_{k} (\zeta - a_k)^{\mu_k}$, $\sum_{k} {\mu_k} = 2s$ (obvious notation),
the Madelung-Bohm velocity form $v^{\chi} = \Im \, d \log \chi \equiv d \varphi$
(with $\varphi$ a local phase function), which is,
in general, 
a closed form on $S^2 \setminus \{Z_i , P_i \}$ (zeros and poles).
In this way, we are looking at  (punctured) $S^2$ as a {\it configuration} space.  Before proceeding, it may be useful to recall the simplest example
of Madelung-Bohm velocity, that for a wave function of a particle of mass one on the real line, $\psi (x) = \mid\!\!{\psi (x)}\!\!\mid \!\!e^{i \phi (x)}$: one has
$v^{\psi} :=  \Im (d \log \psi / dx ) = d\phi/ dx$.
Resuming our discussion,
we have, working on ${\bf C}$,
\begin{equation}
\! d v^{\chi} = 2\pi \sum_{k} {\mu_k}\, \delta_{a_k}(\zeta) \, ({i\over{2}} d\zeta \wedge d{\overline{\zeta} })
= 2\pi \sum_{k} {\mu_k}\, \delta_{a_k}(\zeta) \,  dx \wedge dy,
\quad \delta v^{\chi} = 0
\end{equation}
($\delta$ denotes divergence).
In the r.h.s. we have a {\it singular} ($\delta$-like) vorticity 2-form (viewed as a current (singular Poincar\'e dual),
see \cite{Pe-Spe}, \cite{Sp2}, \cite{BT}, \cite{Gr-Har})
corresponding to the {\it vorticity divisor} 
$D = \sum_{k} {\mu_k} \, {a_k}$, which represents an assembly of point vortices located at $a_k$, with strength
$\mu_k$.
Now, letting ${\gamma}$ be a circuit encirling - once, counterclockwise - some of the roots $a_k$,
and invoking the Residue Theorem, we immediately reach the following conclusion:

\begin{Theorem}
(Spin = Vorticity) (i) With the notation above

\begin{equation}
{1\over {2\pi}}\int_{\gamma} v^{\chi} = \sum {\mu_k}
\end{equation}
where the sum ranges over the roots encircled by $\gamma$.\par
\smallskip
\noindent
(ii)
In particular, if 
the circuit ${\Gamma}$ encircles once,
counterclockwise, all the zeros of $\chi$, then
\begin{equation}
{1\over {2\pi}}\, \int_{\Gamma} v^{\chi} = 2s
\end{equation}
that is,
 the total spin is given by the circulation of the velocity field along a loop encirling
once the zeros of the wave functions, so
ultimately it can be looked upon as a 
(quantized) {\rm vorticity} strength.\par
\smallskip
\noindent
(iii)  The above velocity can be interpreted as a (flat) connection (form) on the trivial complex line bundle over the (punctured) sphere, and fulfils,
for definiteness of the spin wave function (i.e. trivial holonomy)
a {\rm Bohr-Sommerfeld} type quantization condition which is tantamount to  the {\rm Feynman-Onsager} one.
\par
\end{Theorem}

{\bf Remark.}  As we have already noticed, parts (i) and (ii) of the above result are, strictly speaking, just a rephrasal of the residue theorem.
However, the main point is that upon using the above complex polynomial representation of spin wave functions we get the sought for hydrodynamical and configuration space interpretation of spin. Part (iii) is clear after tracing back the relevant definitions, and observing that $H^{1}(S^2 \setminus \{ \hbox{punctures} \} )\cong {\bf Z}^p$, where $p$ denotes the number of punctures (first homology group of the configuration space (with punctures), viewed as a Lagrangian submanifold of cotangent space);  the spin wave function is thus formally viewed as a semiclassical wave function, defined on an appropriate Lagrangian submanifold and subject
to Bohr-Sommerfeld type conditions, see also \cite{Tyu}, \cite{Be-Spe},
 \cite{Sp2}, and references therein;
 see e.g. \cite{Gue} for a physical discussion of the Feynman-Onsager condition.\par

\section{Hydrodynamical properties of Killing vector fields}
In this section we discuss some (possibly new or, at least, differently formulated) results 
valid for Killing vector fields on a (connected) Riemannian manifold $(M,g)$ 
(i.e. those generating infinitesimal isometries; they always exist, at least locally).
As general references we may quote \cite{KN}, \cite{GHL}, \cite{DC}.
For hydrodynamics we refer, among others, to \cite{Arn-Khe}, \cite{Tay}, \cite{Eb-Mar}, 
\cite{Abr-Mar}, \cite{MW}.\par
The Levi-Civita connection of $(M,g)$ will be denoted by $\nabla$.
We shall employ the notation $ \langle X, Y \rangle := g (X, Y)$, for $X$, $Y \in \Gamma (TM)$
(vector fields on $M$). Upon freely using the musical isomorphism notation ($\sharp$ = vector field,
$\flat$ = 1-form, corresponding to index raising and lowering, respectively, 
so, for instance, $(X^{\flat}, Y) = \langle X, Y \rangle$, with $(\cdot,\cdot)$ being the pairing 
between 1-forms and vector fields),
we begin by
recalling the following basic identity (cf. \cite{Abr-Mar}, 5.5.8, p.474, or \cite{Arn-Khe}, Ch.IV, Theorem 1.17, p.202):

\begin{equation}
{\cal L}_Y  Y^{\flat} = (\nabla_Y Y)^{\flat} + {1\over 2} \, d \,\langle Y , Y \rangle
\end{equation}
(${\cal L}$ is the Lie derivative).
The following result is crucial.
\begin{Lemma}
Let $X$ be a Killing vector field on a Riemannian manifold $(M,g)$. Then

\begin{equation}
{\cal L}_X  X^{\flat} = 0
\end{equation}
\end{Lemma}

{\bf Proof.} If $X$ is Killing, then for any vector field $Y$, one has
\begin{equation}
{\cal L}_X  (Y^{\flat}) = ({\cal L}_X  Y)^{\flat} 
\end{equation}
which yields immediately
\begin{equation}
{\cal L}_X  (X^{\flat}) = ({\cal L}_X  X)^{\flat}  = [X, X]^{\flat} = 0
\end{equation}
Q.E.D.\par
\smallskip
Recall that the {\it Euler equation} on a Riemannian manifold reads, among others, in the following
equivalent guises, in terms of 1-forms:
\begin{equation}
{{\partial X^{\flat}} \over {\partial t}} + (\nabla_X X)^{\flat} = - d \, p
\end{equation}
or (cf. (3.1))
\begin{equation}
{{\partial X^{\flat}} \over {\partial t}}\, + \,  {\cal L}_X X^{\flat} = 
d \,( {1\over 2} \langle X, X \rangle \,- \, p)
\end{equation}
($p$ being the {\it pressure}) together with ${\sl div} X = 0$ (see e.g. \cite{Arn-Khe} or \cite{Tay}, Ch.17, 1.15,
p.469). One immediately establishes the following 
\begin{Lemma}
A divergence-free vector field $Y$  on a (finite dimensional, connected) Riemannian
manifold $(M,g)$ satisfies the {\rm stationary Euler equation},
with {\rm pressure} $p = {1\over 2}\langle Y, Y \rangle$ (up to a constant) if and only if ${\cal L}_Y  Y^{\flat} = 0$.
\end{Lemma}
Let us also notice, for future use, the general identity, valid for a Killing vector field $X$,
\begin{equation}
({\cal L}_X g )( Y, Z ) \, = \,\langle \nabla_Y X , Z \rangle + \langle \nabla_Z X , Y \rangle = 0
\end{equation}
which implies, setting $Z = Y$,  
\begin{equation}
\langle   Y , \nabla_Y X\rangle  = 0
\end{equation}
and, setting further $Y = X$,
\begin{equation}
\langle   X , \nabla_X  X\rangle  = 0 \, .
\end{equation}
The main result of this section is the following
\begin{Theorem}
Let $X$ be a  Killing vector field on a finite dimensional, connected Riemannian
manifold $(M,g)$. Then:\par
\noindent
(i) the (necessarily divergence-free) vector field  $X$ fulfils the {\rm stationary Euler equation}, with {\rm pressure} given by 
$p = {1\over 2}\langle X , X \rangle$ (up to a constant);\par
\smallskip
\noindent
(ii) the {\rm vorticity} form of the (stationary) Euler equation reads (with $w = d X^{\flat}$ the {\rm vorticity}
2-form)
\begin{equation}
{\cal L}_X w = 0 \, ;
\end{equation}
\smallskip
\noindent
(iii) the (Riemannian) gradient of the pressure, $ (dp)^{\sharp}$, is orthogonal to $X$;\par
\smallskip
\noindent
(iv) if $\gamma$ is an integral curve of $X$ starting from a point $m \in M$,
then $\gamma$ is a geodesic if and only if $dp = 0$ (at $m$ and hence along $\gamma$).\par

\end{Theorem}

{\bf Proof.}
Ad (i).
The  conclusion follows immediately from Lemmata 3.1 and 3.2.\par
\noindent
Ad (ii). This is clear from (3.2) and the fact that ${\cal L} \, d = d \, {\cal L}$.\par
\smallskip
\noindent
Ad (iii). This is straightforward from $ (dp)^{\sharp} = - \nabla_X X $ 
and from (3.9). \par
\smallskip
\noindent
Ad (iv). Let $\gamma;:s \mapsto \gamma(s)$ denote the integral curve of $X$ starting from a point $m$. Then, due to the stationary Euler equation fulfilled by
$X$, one has
\begin{equation}
(\nabla_{\dot{\gamma}} {\dot{\gamma}})^{\flat} = - d p \mid_{\gamma(s)}
\end{equation}
Thus $\gamma$ is a geodesic if and only if  $d p \mid_{\gamma(s)} = 0$ for all $s$. On the other hand, $p$, and hence $dp$ are invariant under the flow
of $X$, by (iii), whence $d p \mid_{\gamma(s)} = 0$ for all $s$ if and only if it holds at $m = \gamma(0)$, this yielding (iv).\par

\smallskip
Let us also recall and prove, for completeness, the following
\begin{Proposition}
(cf. \cite{Tay}). Along a geodesic $\gamma$, if the vector field \ $Y$ restricts to its velocity field thereon, and
$X$ is Killing, we have 
\begin{equation}
Y \langle  Y , X \rangle = 0
\end{equation}
\end{Proposition}

{\bf Proof.} One has, along $\gamma$
\begin{equation}
Y \langle Y, X \rangle \, = \, \langle \nabla_Y Y , X \rangle + \langle   Y , \nabla_Y X\rangle 
= \langle   Y , \nabla_Y X\rangle . 
\end{equation}
and, by (3.8), the conclusion.
Q.E.D.\par
\smallskip
{\bf Remarks.} 1. The above proposition says that the scalar product of the velocity field of a geodesic
with a Killing field is conserved (see e.g. \cite{Tay}, Ch.18, proposition 3.3, p.546).
For surfaces of revolution, it amounts to the classical Clairaut's Theorem.\par
\smallskip
\noindent
2. Assertion (iii) of Theorem 3.1 also appears in \cite{KN}, Prop. 5.7, p.252
(one direction, and the proof is different) and can be also proved by exploiting the variational characterization
of geodesics as critical paths of the energy functional (together with the Killing condition).
Application to a surface of revolution (along the $z$-axis, say) yields the standard 
characterization of geodesic
parallels as extremals of the radial function (i.e. the profile curve viewed as a function of $z$ ):
indeed, using standard notation, one has $X = {{\partial} \over{ \partial \varphi}}$ and  
$g({{\partial} \over{ \partial \varphi}}, {{\partial} \over{ \partial \varphi}}) = \varrho^2$,
where $ g = (1+ {\varrho^{\prime}}^2 ) dz^2 + \varrho^2 d \varphi^2$.\par
\smallskip
\noindent
3. Notice that for one-sided invariant metrics on Lie groups, even in the finite dimensional case,
geodesics do not correspond to 1-parameter Lie subgroups
(see e.g. \cite{GHL}) so, even ignoring the subtleties of
the infinite dimensional situation, one cannot directly conclude that a divergence-free
vector field on a (compact, say) Riemannian manifold 
(i.e. an element of the ``Lie algebra" of the group $Sdiff(M)$ of measure preserving diffeomorphisms
of $M$) automatically yields a solution of the (stationary)
Euler equation (i.e. a geodesic of the natural right-invariant (but not bi-invariant) metric induced by the
kinetic energy (\cite{Arn-Khe}, \cite{Eb-Mar}). Thus, our specific observations on Killing vector fields
might be useful.\par

\section{A quantum mechanical application}
In this section we wish to apply the results of the preceding section to the velocity vector
field determined by the Schr\"odinger equation, for time independent Hamiltonians and in finite
dimensional quantum Hilbert spaces (which is not so severe a limitation, in view of their occurence 
in various contexts, from quantum chemistry to quantum computing), and we begin by
reviewing briefly the formalism of geometric quantum mechanics,
referring to \cite{Sp1}, \cite{Be-Sa-Spe}, \cite{Ben-Spe} for notation and full details
(but see also \cite{Ash-Sch}, \cite{Bro-Hug}, \cite{Cir-Gat-Man}, \cite{Cir-Man-Piz}, \cite{Cir-Piz}, \cite{CJ}).
This will provide a genuine higher (even) dimensional example of perfect fluid.
We assume $\hbar = 1$.
Let $V$ be a complex Hilbert space of finite dimension $n+1$,
with scalar product $\langle \cdot | \cdot \rangle$, linear in the second variable.
Let $P(V) \cong {\bf P}({\bf C}^{n+1}) \equiv {\bf P}^n $ denote
its associated projective space, of complex dimension $n$. This is the space
of (pure) states in quantum mechanics. Upon free employ of Dirac's bra-ket
notation, we can identify a point in $P(V)$, which is, by definition,
the ray (i.e. one-dimensional vector space) $<v>$ 
pertaining to (resp. generated by) a non zero vector $ v \equiv |v\rangle$ - and
often conveniently denoted by $[v]$- with
the projection operator onto that line, namely 
\begin{equation}
[v] ={{| v \rangle \langle v  |} \over {\,\,{\parallel v \parallel}^2} }
\end{equation}  
(actually, the above identification can be interpreted in terms of 
a  moment map, see \cite{Be-Sa-Spe}).
If $U(V)$ denotes the unitary group pertaining to $V$, with Lie algebra
$\sl{u}(V)$, consisting of all skew-hermitian endomorphisms of $V$
- which we call observables, with a slight abuse of language - then
the projective space $P(V)$ is a $U(V)$-homogeneous K\"ahler manifold.
 The isotropy group (stabilizer)  of a point $[v] \in P(V)$ 
 is isomorphic to $U(V^{\prime}) \times U(1)$,
with $V^{\prime}$ the orthogonal complement of $<v>$ in $V$, the  $U(1)$ part
coming from phase invariance: $[e^{i\alpha}v] = [v] $.  
Hence
\begin{equation}
P(V) \cong  U(V)/(U(V^{\prime}) \times U(1)) \cong U(n+1)/(U(n) \times U(1))
\end{equation} 
The fundamental vector field $A^{\sharp}$ associated to $A \in \sl{u}(V)$
reads (evaluated at $[v] \in P(V)$, $\parallel v \parallel = 1$)
\begin{equation}
A^{\sharp}|_{[v]} = | v \rangle \langle A v  | + | A v \rangle \langle  v  |
\end{equation}
One finds, for the {\it dispersion} (or variance, or uncertainty) squared of the observable $A$ in the state $[v]$:
\begin{equation}
(\Delta_{[v]} A)^2 := \parallel A v - \langle A v  | v \rangle v\parallel^2 = g_{FS} (A^{\sharp}, A^{\sharp})
\end{equation}
 with $g_{FS}$ the Fubini-Study metric on $P(V)$ (see the references given above).
 We deal with a non degenerate Hamiltonian ($\lambda_i < \lambda_j$ for $i < j$)
\begin{equation}
H = \sum_{i=0}^{n} \lambda_i | e_i \rangle \langle e_i |
\end{equation}
(in terms of an orthonormal basis $(e_i), \, i = 0,...,n$ of $V$).
We write, for a generic state vector (of norm one)
\begin{equation}
v = \sum_{i=0}^{n} \alpha_i e_i , \qquad \sum_{i=0}^{n} |\alpha_i|^2 = 1 .
\end{equation}
The dispersion (squared) of the hamiltonian $H$ in the state $[v]$ is easily computed:
\begin{equation}
(\Delta H)^2 := (\Delta_{[v]} H)^2 = \langle v | H^2 v \rangle  \, - \, \langle v| H v \rangle^2 =  
g_{FS} ((-i H)^{\sharp}, (-i H)^{\sharp})
\end{equation}

The vector field $X =: (-i H)^{\sharp}$ is called the Schr\"odinger vector field on ${\bf P}^n $
(the Schr\"odinger equation reads, of course, $\partial_t \!\mid \!v \rangle = -i H \!\mid \! v \rangle$)
and is Killing thereon (hence divergence-free).
It is also stationary since the Hamiltonian $H$ is time independent.
\par
We shall use the representation ${\bf P}^n \equiv {\bf P}({\bf C}^{n+1}) \cong S^{2n+1}/ S^{1}$,
where $S^{2n+1}$ is the $2n+1$-dimensional sphere in ${\bf C}^{n+1} \cong {\bf R}^{2(n+1)}$.\par
Then Theorem 3.1 immediately implies part
of the following
\begin{Theorem}
(i) If $(M,g) = ({\bf P}^n, g_{FS})$, and  $X$ is the Schr\"odinger vector field
pertaining to the Hamiltonian $H$, then $X$ fulfils the {\rm stationary Euler equation} with $2 p = (\Delta H)^2$.\par
\smallskip
\noindent
(ii) The critical points of the {\rm pressure}, in the Schr\"odinger case, are given by the energy eigenstates
(minima, zero pressure) and by the equal probability superpositions of pairs thereof. \par
\smallskip
\noindent
(iii) The {\rm vorticity} 2-form $w = dX^{\flat}$, evaluated on the geodesic sphere $S_{ij}$ -
with area 2-form $d\sigma$ and colatitude $\vartheta$ - determined by the
superpositions of two energy eigenstates, reads (see below for details):
\begin{equation}
w|_{S_{ij}} = 2 \,(\delta h)_{ij} \,\cos \vartheta\, d \sigma \, .
\end{equation}
\end{Theorem}
{\bf Proof.} Ad (i). This is just an application of Theorem 3.1, (i). Of course, the remaining assertions
of that result hold in the present case.
As a consistency check (see also the third remark in the preceding section) observe that, 
in the projective line (Riemann sphere) case, on the equator
one has critical (actually maximal) uncertainty and the Schr\"odinger trajectory is a geodesic.\par
\smallskip
\noindent
Ad (ii). In order to determine the critical points of the quantum mechanical pressure field explicitly,
we proceed as follows.\par
Set $\varrho_i^2 = |\alpha_i|^2$ and $f = (\Delta H)^2 $ as a function of the $\varrho_i$,
namely

\begin{equation}
f = \sum_{i=0}^{n} \lambda_i^2 \varrho_i^2 - (\sum_{i=0}^{n} \lambda_i \varrho_i^2 )^2
\end{equation}
and introduce the constraint $g = \sum_{i=0}^{n} \varrho_i^2 \, \,- \,1 = 0$.
Then the critical points of $f$, subject to $g=0$, are given by the solutions of
the (Lagrange) system
\begin{equation}
df = \mu \, dg, \qquad g=0
\end{equation}
namely
\begin{equation}
(\lambda_i^2  - 2\, \langle v H v \rangle\, \lambda_i - \mu) \varrho_i = 0\qquad \forall i=0,...,n
\end{equation}
Upon defining $P(\lambda ) = \lambda^2 - 2 \langle v | H v \rangle - \mu$, we see that, if 
we have a solution with $\varrho_k \neq 0$,
then $\lambda_k$ must be a root of $P$.
Therefore, since the eigenvalues are all distinct,
there are at most two indices $i_1$, $i_2$ for which $\varrho_i \neq 0$, and this leads
to $\langle v | H v \rangle = \lambda_{i_1} \varrho_{i_1}^2 +  \lambda_{i_2} \varrho_{i_2}^2  = 
{1\over 2}(\lambda_{i_1}  +  \lambda_{i_2}) $, whencefrom it follows that $\varrho_{i_1}^2 = \varrho_{i_2}^2 = {1\over 2}$,
and $\mu =  - \lambda_{i_1}  \,  \lambda_{i_2} $.
The remaining possibility, that only one $\varrho_i \neq 0$, yields the eigenstates of $H$.  \par
\smallskip
\noindent
Ad (iii).
In computing the vorticity 2-form $d X^{\flat}$ pertaining to the Schr\"odinger velocity 1-form $X^{\flat}$,
we first notice that
in view of the previous
discussion, it is enough, in order to grasp its physical meaning, to restrict to the (totally) geodesic spheres 
${S}_{ij}$, say, 
determined by superpositions of two
energy eigenstates. The Schr\"odinger motion is just a uniform rotation around the axis whose poles are
given by the eigenstates in question (see also \cite{Ben-Spe}, \cite{Be-Sa-Spe}); the angular velocity $\omega
\equiv (\delta h)_{ij}$ equals $\lambda_i - \lambda_j$ ($i > j$), 
the difference of the energy levels. 
We find ($\vartheta$ is the colatitude, measured appropriately, and $d \sigma $ is area 2-form; also recall that the radius
$R = {1\over 2}$, cf. \cite{Ben-Spe}, \cite{CJ}):
\begin{equation}
w|_{S_{ij}} = d X^{\flat}|_{S_{ij}} = d (\omega \, \,R^2 \,\sin^2 \vartheta d\varphi) = 2 \,\omega\, \cos \vartheta \,(R^2 \sin \vartheta 
\, d\vartheta \wedge d\varphi)
= 2 \,\omega \,\cos \vartheta\, d \sigma
\end{equation}
\noindent
whence the vorticity vanishes on the equator (maximal uncertainty) and it is maximal (with opposite signs) at the poles
(zero uncertainty).  Notice, as a further check, that the {\it scalar} vorticity function $\widetilde{w} :=  2 \,\omega \,\cos \vartheta$ does indeed satisfy the
2d-vorticity equation  on $S_{ij}$ (obvious notation, cf. \cite{Tay}, Ch.17, (1.27) p.470)
\begin{equation}
\frac{\partial \widetilde{w} }{\partial t} + {\sl grad} \, \widetilde{w} \cdot X = 0
\end{equation}
\ Q.E.D.\par
\smallskip
\noindent

{\bf Remarks.}
1. In geometric terms, the critical points
are given by the vertices and the midpoints of the Atiyah - Guillemin-Sternberg convex polytope arising
from the standard moment map (cf. \cite{Be-Sa-Spe}, \cite{McD-Sal} for background).\par
\smallskip
\noindent
2. In essence, we provided an ``Eulerian" counterpart to the ``Lagrangian" portrait inherent
to the geometric interpretation of the Schr\"odinger flow.\par
\smallskip
\noindent
3. We may depict the following picture of the ``collapse of the wave function":
performing an energy measurement on a quantum system causes a perturbation of the Schr\"odinger fluid,
forcing the quantum state to reach to a minimal (indeed, zero) pressure, i.e. an eigenstate
(see also \cite{Be-Sa-Spe} for a complementary discussion of this issue).\par
\smallskip
\noindent
4. The geometrical and hydrodynamical  set up  may be useful in ``visualising" the Quantum Zeno Effect (see e.g. \cite{JZKGKS}, 3.3.1, p.110): continual measurement ``freezes" the
motion: the rate of decay of a pure state (as a function of  $t$) goes as $(\Delta H)^2 t^2$, the  ``space" (squared) travelled by the state under the Schr\"odinger
motion (Lagrangian portrait), and related in turn to the fluid pressure. Upon repeating the measurement $N$ times within the time
interval $t$ one finds $(\Delta H)^2 \frac{t^2}{N}$, tending to zero as $N$ goes to infinity.\par
\smallskip
\noindent
5. Let us remark on the similarity between the general geometric quantum mechanical picture  and that of an assembly of  harmonic oscillators (also cf. \cite{H} and the general discussion about integrability in \cite{Be-Sa-Spe}. Indeed,  projective space comes from a Marsden-Weinstein reduction of the phase space of latter (see e.g. \cite{Ga-Spe}). Following this path one again arrives at the conclusion that the Schr\"odinger field fulfils the stationary Euler equation. However, the general argument we gave is by no means more complicated and it is more intrinsic, allowing the extra consequences of Theorem 4.1. to be drawn.\par

\section{Concluding remarks}

We close the present note with the following additional observations.\par
\smallskip
\noindent
1. The geodesic interpretation of Euler's equation
entails that the Schr\"odinger equation
can be viewed as coming from a hydrodynamical variational
principle in projective space, equipped with the Fubini-Study volume, via the Killing condition
(yielding the natural $U(n+1)$-symmetry of ${\bf P}^n$).\par
\smallskip
\noindent
2. Notice that the Schr\"odinger motion itself can be viewed as a coadjoint orbit motion
for the group $U(n+1)$ (see e.g. \cite{Be-Sa-Spe} for full details). On the other hand,
 the vorticity form of the
Euler equation is a manifestation of a coadjoint orbit motion relative to the group of measure
preserving diffeomorphisms (\cite{Arn-Khe}, \cite{MW}). In our case we deal with a stationary fluid, and we arrive at equation (3.10).\par
\smallskip
\noindent
3. Since the quantum state space $ ({\bf P}^n, g_{FS})$ is a K\"ahler-Einstein manifold (the ``cosmological" constant is indeed a pressure term) 
the Schr\"odinger equation appears to be a (Killing) symmetry for a fictitious (Riemannian)
``general relativity" (cf. Proposition 3.1) thereon, ultimately governed by uncertainty.\par

\bigskip

{\bf Acknowledgements.} The author is grateful to F. Cardin, L.M. Morato and N. Sansonetto for
useful comments. He is indebted to MIUR (ex 60\%) for financial support. \par

\end{document}